\documentclass[twocolumn,showpacs,prb]{revtex4}
\usepackage{graphicx}

\begin{document}

\title{
Para-ortho transition of artificial H$_2$ molecule in lateral   
quantum dots doped with magnetic impurities
}

\author{Ramin M. Abolfath}

\affiliation{
Department of Radiation Oncology, University of Texas Southwestern
Medical Center, Dallas, Texas 75390, USA
}

\date{\today}

\begin{abstract}
We present the magnetic phase diagram of artificial H$_2$ molecule 
in lateral quantum dots doped with magnetic impurities as a function
of external magnetic field and plunger gate voltage.
The onset of Mn-Mn antiferromagnetic-ferromagnetic transition 
follows the electron spin singlet-triplet transition.
We deploy a configuration-interaction method to exactly diagonalize
the electron-Mn Hamiltonian
and map it to an effective Mn-Mn Heisenberg Hamiltonian.
We find that Mn-Mn exchange coupling can be described by 
RKKY-interaction/magnetic-polaron in weak/strong electron-Mn coupling 
at low/high magnetic fields.
\end{abstract}

\maketitle

\section{Introduction}
There is currently significant experimental 
\cite{Jacak:book,Besombes2004:PRL,Gould2006:PRL,Mackowski2008:PRL,Bussian2008:NM,Bree2008:PRB,Reiter2009:PRL}
and theoretical 
\cite{Fernandez2004:PRL,Govorov2005:PRB,Qu2006:PRL,Abolfath2008:PRL,Cheng2009:PRB,Nguyen2008:PRB,Arrondo2009:PRB}
interest in semiconductor quantum dots (QDs) doped with magnetic impurities.
Fabrication of hybrid systems consisting magnetic ions in a controlled
electronic environment provides an interesting interplay between interaction
effects and magnetism.
In particular the application of spin of electrons in quantum dot 
molecules for generation of electron entanglement and quantum information
processing in solid state devices is of current interest 
\cite{QDMexperiment,RaminWojtekPawel}.
One of the challenges in use of spin of electrons in scalable 
quantum computer devices is the spin dephasing
due to interference by spin-orbit coupling and nuclear hyperfine interaction
\cite{Cywiski2009:PRL}.
The broadening of the electron envelop wavefunction in a 
QD is determined by the length scale of the confining potential 
that is comparable with the size of QD (typically $\approx$ 1-1000 nm).
In materials with high abundance of spin-carrying nuclear isotopes,
the electron interacts with large number of nuclei in host semiconductor. 
Although the strength of nuclear hyperfine coupling is small 
(typically $\approx 1\mu$eV) compare to
other relevant energy scales, but because of the broadening of 
the confined electron in nano-meter length scale, it interacts
with a large number of nuclei that contribute to the electron spin dephasing.
The use of magnetic moment of nuclear impurities in 
host semiconductors for the quantum information processing
have appeared to overcome this limitation
\cite{Cory1998:Nature,Kane:1998:Nature}.
Recently the system of 
$^{13}$C atoms in two-electron nanotube quantum dot molecules has 
been studied \cite{Churchill2009:PRL}.
The advantage of using singlet eigen-states of 
coupled magnetic moments of nucleus of $^{13}$C 
in organic molecules and their effects
in dramatic enhancement of the spin life-times needed for imaging the 
metabolic pathways in living systems by hyperpolarization 
methods has been investigated~\cite{Warren2009:Science}.
Moreover, molecules doped with $^{13}$C has been used in
demonstration of quantum teleportation using NMR techniques
\cite{Cory1998:Nature}.  
Similarly the magnetic dipole moment of nucleus of
$^{31}$P-impurities in Si-based quantum computer model  
proposed by Kane~\cite{Kane:1998:Nature} 
appears to be promising in making quantum computer 
solid state devices.
Similar to magnetic dipole moment of nucleus of $^{13}$C and/or $^{31}$P,
the magnetic moment of electrons in $d$-shell  
of magnetic impurities (such as Mn, Fe, Co)
in lateral quantum dot molecules can be used 
for quantum information processing
as they have been used for fabrication of
molecular magnets~\cite{Almeida2009:PRB}.

In this work we focus on theoretical study of the phase diagram and spin 
transitions of coupled magnetic impurities (e.g. Mn)
doped in two electron quantum dot molecule, an artificial
H$_2$ molecule.  
Here we calculate the Mn-Mn effective Hamiltonian mediated by electrons 
and show that the  
onset of spin-polarized state with finite Mn-magnetization, 
corresponds to spin singlet-triplet transition of two electrons in QD molecule.
This transition is analogous to para-ortho transition of nucleus of solid
H$_2$ where the electron-nuclei hyperfine interaction opens the energy gap
between para and ortho states of H$_2$ 
molecule~\cite{LandauLifshitzQM:book,comment}.
In the small (large) magnetic fields the spin singlet (triplet) 
state is the ground state of electrons \cite{RaminWojtekPawel}
and thus the ground state of 
the coupled Mn is described by hydrogen molecule para (ortho) state.
In the small magnetic fields the Mn-Mn interaction induced by electrons 
is calculated perturbatively in the electron-Mn weak interacting limit.
It can be described effectively by RKKY-coupling  \cite{RKKY}. 
In large magnetic field the ground state of electrons is spin-triplet.
The electron-Mn interaction is strong and magnetic-polaron state form.
In this limit the ortho-state of artificial H$_2$ molecule is stable.
We show the Mn-Mn exchange coupling can be controlled by inter and intra dot
correlations, external magnetic field and gate voltage.
The dependence of Mn-Mn interaction to 
the external gate voltage and magnetic field mediated by
spin singlet-triplet transition
among electrons in QDs opens up the possibility in using Mn-magnetic moment as
qubit in quantum computation purposes.
In contrast to electrons confined in QDs, 
because of highly localized d-electrons, Mn's interact directly with 
significantly smaller number of nuclei in host semiconductor. 
The Mn d-electrons also do not interact directly with 
the host semiconductor Rashba and Dresselhaus 
spin-orbit coupling~\cite{Dresselhaus1955:PR}, 
and thus their spin coherence life-time is expected to be longer than the
QD electrons.
However, Mn-Mn interaction mediated by electrons in QDs are still vulnerable 
to the electron spin dephasing mediated by QD electrons due to their hyperfine 
and spin-orbit couplings. 
In analogous to the $^{31}$P in Si system~\cite{Kane:1998:Nature}, 
the spin coherence time in the Mn system is expected to be longer than
the QD electron system.
Further investigation is required to make a quantitative dependence
of spin decoherence time of Mn on the electron-nuclear hyperfine interaction
and their spin-orbit coupling.

\section{Hamiltonian}
We represent magnetic QD molecule by the
Hamiltonian $H=H_e+H_{em}+H_m$, describing contributions of
interacting electrons, electron-Mn (e-Mn) exchange, and
direct Mn-Mn antiferromagnetic (AFM) coupling, respectively.
Electrons confined in quasi-two-dimensional 
quantum dots in a uniform perpendicular 
magnetic field can be described by the effective mass Hamiltonian
\begin{eqnarray}
H_e = \sum_{i=1}^N  \left( T_i + E_{iZ} \right)
+ \frac{e^2}{2\epsilon}\sum_{i \neq j}\frac{1}{|\vec{r}_i - \vec{r}_j|},
\label{ham1}
\end{eqnarray}
where
\begin{eqnarray}
T=\frac{1}{2m^*}\left(\frac{\hbar}{i}\vec{\nabla}
+ \frac{e}{c} A(\vec{r})\right)^2 + V(x,y)
\label{ham2}
\end{eqnarray}
is the single electron Hamiltonian in magnetic field. Here $(\vec{r})=(x,y)$
describes electron position, $V(\vec{r})$ is the quantum dots confining 
potential, $A(\vec{r})=\frac{1}{2}\vec{B}\times\vec{r}$ 
is the vector potential, and $B$ is the external magnetic field perpendicular
to the plane of confining potential.
$m^*$ is the conduction-electron effective mass, 
$-e$ is the electron charge, and
$\epsilon$ is the host semiconductor dielectric constant.
$E_{iZ}= \frac{1}{2}g_e\mu_B \sigma_{iz} B$ is the Zeeman spin splitting, 
$g_e$ is the electron g-factor in host semiconductor,
$\mu_B$ is the Bohr magneton, and $\sigma$ is the Pauli matrix.
The single particle eigenvalues ($\epsilon_{\alpha\sigma}$) and 
eigenvectors ($\varphi_{\alpha\sigma}$) 
are calculated by
discretizing $T$ in real space, and diagonalizing the resulting matrix
using conjugated gradient algorithm \cite{RaminWojtekPawel,JCP06}.
The single-particle (SP) states can be used as a basis in
configuration-interaction (CI) calculation that allows to diagonalize
Hamiltonian $H$.
The details of CI method can be found in Ref.~\cite{JCP06}.
Denoting the creation (annihilation) operators for electron in
non-interacting SP state $|\alpha\sigma\rangle$ by 
$c^\dagger_{\alpha\sigma}~(c_{\alpha\sigma})$,
the Hamiltonian of an interacting system in second quantization
can be written as
\begin{eqnarray}
H_e&=&\sum_{\alpha}\sum_{\sigma} 
\epsilon_{\alpha\sigma}
c^\dagger_{\alpha\sigma} c_{\alpha\sigma} \nonumber \\ &&
+ \frac{1}{2}\sum_{\alpha\beta\gamma\mu}\sum_{\sigma\sigma'}
V_{\alpha\sigma,\beta\sigma',\gamma\sigma',\mu\sigma}
c^\dagger_{\alpha\sigma} c^\dagger_{\beta\sigma'} 
c_{\gamma\sigma'} c_{\mu\sigma},
\label{multiparticle}
\end{eqnarray}       
where the first term represents the single particle Hamiltonian 
Eq.(\ref{ham2}), 
and
${V}_{\alpha\sigma,\beta\sigma',\mu\sigma',\nu\sigma} = 
\int d\vec{r} \int d\vec{r'}  
{\varphi}^*_{\alpha\sigma}(\vec{r})
{\varphi}^*_{\beta\sigma'}(\vec{r'}) 
\frac{e^2}{\epsilon|\vec{r}-\vec{r'}|}
{\varphi}_{\mu\sigma'}(\vec{r'})
{\varphi}_{\nu\sigma}(\vec{r})$, 
is the two-body Coulomb matrix element.
We describe e-Mn exchange interaction by
$H_{em}=- J_{sd}\sum_{i,I}\vec{s}_i\cdot\vec{M}_I  
\delta({\bf r}_i - {\bf R}_I)$,
where $J_{sd}$ is the exchange coupling between electron spin 
$\vec{s}_i$ at ${\bf r}_i$
and impurity spin $\vec{M}_I$ at ${\bf R}_I$~\cite{Qu2006:PRL,Abolfath2008:PRL}.
Note that $H_{em}$ is analogous of the isotropic (contact) part of electron-nucleus hyperfine interaction~\cite{coish2009:PSSB},
responsible for para-ortho energy gap of solid $H_2$ (for comparison 
see for example eq. 121.9, page 498 of Ref.~\cite{LandauLifshitzQM:book}).
In second quantization it can be written as
\begin{eqnarray}
H_{em} &=& - \sum_{\alpha\beta}\sum_I \frac{J_{\alpha\beta}({\bf R}_I)}{2} 
[M_{zI}
(c^\dagger_{\alpha\uparrow}c_{\beta\uparrow}-
c^\dagger_{\alpha\downarrow}c_{\beta\downarrow}) \nonumber \\ &&
+ M_I^+ c^\dagger_{\alpha\downarrow}c_{\beta\uparrow}
+ M_I^- c^\dagger_{\alpha\uparrow}c_{\beta\downarrow}],
\label{Hex}
\end{eqnarray}       
where $J_{\alpha\beta}({\bf R}_I) = J_{sd}\varphi^*_{\alpha}({\bf R}_I)
\varphi_{\beta}({\bf R}_I)$.
Finally we describe Mn-Mn direct exchange interaction and 
Mn-Zeeman coupling by
$H_m=\sum_{I,I'} J^{AF}_{I,I'} \vec{M}_I \cdot \vec{M}_{I'}
+ \sum_{I} g_m \mu_B M_{Iz} B$, where 
$J^{AF}_{I,I'}$ 
is the direct Mn-Mn AFM coupling~\cite{Qu2006:PRL}, resembling the 
direct dipole-dipole interaction, and $g_m$ is the Mn g-factor.

\subsection{Confining potential}
For numerical calculation we model quantum dots molecules by the
following confining potential 
$V(x,y)=V_L~ \exp[{-\frac{(x+a)^2+y^2}{\Delta^2}}]
       +V_R~ \exp[{-\frac{(x-a)^2+y^2}{\Delta^2}}]
+V_p \exp[{-\frac{x^2}{\Delta_{Px}^2}-\frac{y^2}{\Delta_{Py}^2}}]$.
Here $V_L,V_R$ describe the depth of the left and right quantum dot minima
located at $x=-a,y=0$ and $x=+a,y=0$, and $V_p$ is the plunger gate potential
controlled by the central gate~\cite{JCP06}. 
For identical dots, $V_L=V_R=V_0$, 
and confining potential  exhibits inversion symmetry. 
Our numerical results are calculated for parameters 
based on (Cd,Mn)Te  QDs 
with $J_{sd}=0.015$ eV nm$^3$, $m^*=0.106$, 
$\epsilon=10.6$, $g_m=2.02$, $g_e=-1.67$
and the effective Bohr radius $a^*_B=5.29$ nm
and Rydberg energy $Ry^*=12.8$ meV.
We parametrize the confining potential by $V_0=-10, a=2, \Delta=2.5$, 
and $\Delta_{Px}=0.3$, $\Delta_{Py}=2.5$, in effective atomic units. 
$V_p$, which controls the potential barrier,
varies to control the inter-dot correlations,
independent of the locations of the quantum dots.
The choice of parameters ensures weakly coupled quantum dots.

\section{Two level system}
For the purpose of this work we consider a coupled quantum dot system filled
with two electrons. 
It is convenient to project the Hilbert space of two electrons 
into a two level system. 
The construction of two level system based on single particle orbitals
localized in each dot is made by defining a pair of 
bonding-anti-bonding single particle orbitals
$\varphi_\pm(\vec{r})=
\left[\varphi_L(\vec{r}) \pm \varphi_R(\vec{r})\right]/\sqrt{2(1\pm W)}$,
where $\varphi_{L(R)}(\vec{r})$ is the spatial part of SP wave-function
localized in L (R) dot, and $W=Re(\langle L|R\rangle)$ is the overlap 
integral.
At zero (finite) magnetic field the SP orbitals are real (complex) functions.
At $H_{em}=B=0$ the lowest energy many body wave function (ground state)
of two electrons is spatially symmetric with parity +1. 
Thus spin state of the ground state must be singlet:
$\Psi_G(\vec{r}_1, \vec{r}_2) = 
\left[\alpha\varphi_+(\vec{r}_1) \varphi_+(\vec{r}_2)
     + \beta\varphi_-(\vec{r}_1) \varphi_-(\vec{r}_2) \right] |S_0\rangle$,
where $\beta=\sqrt{1-\alpha^2}$ and
$|S_0\rangle = (|\uparrow\downarrow\rangle 
- |\downarrow\uparrow\rangle)/\sqrt{2}$ corresponding to $S=S_z=0$.
Here $S$ is the total spin of two electrons.
The lowest energy excited states are the spin-triplet states with
spatially antisymmetric wave-function corresponding to parity -1, and 
$\Psi^\sigma_X(\vec{r}_1, \vec{r}_2) = \frac{1}{\sqrt{2}} 
\left[\varphi_+(\vec{r}_1) \varphi_-(\vec{r}_2)
     -\varphi_+(\vec{r}_2) \varphi_-(\vec{r}_1)\right] 
     |T_\sigma\rangle$.
Here $\sigma=0,\pm 1$, and $|T_\sigma\rangle$ is one of the spin-triplet states:
$|T_{+1}\rangle = |\uparrow\uparrow\rangle$,
$|T_0\rangle = (|\uparrow\downarrow\rangle 
+ |\downarrow\uparrow\rangle)/\sqrt{2}$, and
$|T_{-1}\rangle = |\downarrow\downarrow\rangle$
corresponding to $S=1$, and $S_z=+1,0,-1$ respectively.
Note that at $B=0$ (or zero Zeeman coupling)
these states are degenerate and thus 
$\Psi^\sigma_X$ is three-fold degenerate.
We define spin singlet-triplet energy gap of electrons 
$\Delta^{(e)}_{ST} \equiv E^{(e)}_X - E^{(e)}_G$ where
$H_e |\Psi_G\rangle = E^{(e)}_G |\Psi_G\rangle$, and
$H_e |\Psi^\sigma_X\rangle = E^{(e)}_X |\Psi^\sigma_X\rangle$. 
In this two level system there are two other excited states with spin-singlet
$\Psi^{s1}_X(\vec{r}_1, \vec{r}_2) = 
\left[\beta\varphi_+(\vec{r}_1) \varphi_+(\vec{r}_2)
    - \alpha\varphi_-(\vec{r}_1) \varphi_-(\vec{r}_2) \right] |S_0\rangle$,
and
$\Psi^{s2}_X(\vec{r}_1, \vec{r}_2) = \frac{1}{\sqrt{2}} 
\left[\varphi_+(\vec{r}_1) \varphi_-(\vec{r}_2)
     +\varphi_+(\vec{r}_2) \varphi_-(\vec{r}_1)\right] 
     |S_0\rangle$.

\begin{figure}
\begin{center}
\vspace{1cm}
\includegraphics[width=0.98\linewidth]{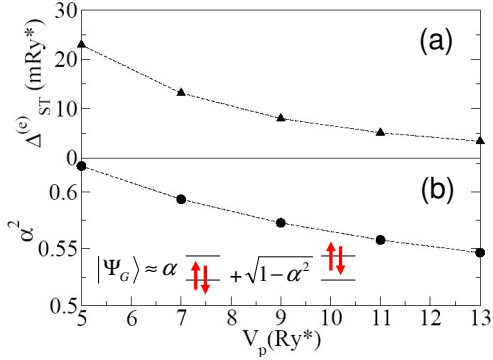}
\caption{
(Color online) 
(a) Shown singlet-triplet energy gap ($\Delta^{(e)}_{ST}$) 
of two electrons in a coupled quantum dot
molecule as a function of plunger gate voltage ($V_p$) at $B=0$.
(b) Shown $\alpha^2$ as a function of $V_p$.
}
\label{fig2}
\end{center}
\end{figure}

\begin{figure}
\begin{center}
\vspace{1cm}
\includegraphics[width=0.98\linewidth]{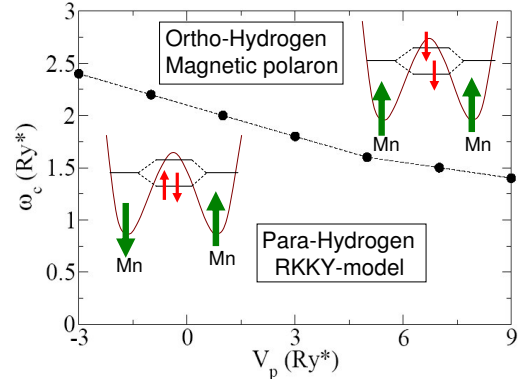}
\caption{
(Color online) 
Spin singlet-triplet phase diagram calculated by 
configuration interaction method for two electrons in lateral quantum
dot molecules. 
}
\label{fig1}
\end{center}
\end{figure}

\subsection{Weak e-Mn coupling}
Assuming the weak electron-Mn interaction limit, we calculate Mn-Mn
effective Hamiltonian mediated by electrons perturbatively. 
In this limit $H_{em}$ is assumed to be small, compared to   
unperturbed Hamiltonian $H_e$. 
In the low magnetic field limit, the ground state of two electrons
in coupled quantum dot is spin-singlet.
It follows
\begin{eqnarray}
H^{\rm eff}_{mm} = \sum_{X} 
\frac{|\langle \Psi_X| H_{em}|\Psi_G\rangle|^2}{E^{(e)}_G - E^{(e)}_X}. 
\label{Heff}
\end{eqnarray} 
Here $G$ and $X$ denote the ground and excited states of electrons in quantum
dot systems, and $X\in \{\Psi^{s1}_X, \Psi^{s2}_X, \Psi^{0,\pm 1}_X\}$.
The uniqueness (non-degeneracy) of the ground state has been assumed implicitly.
To obtain effective interaction between two Mn, we 
calculate the matrix elements of $H_{em}$.
It is straightforward to show that
$\langle \Psi^{s1}_X| H_{em}|\Psi_G\rangle = 
\langle \Psi^{s2}_X| H_{em}|\Psi_G\rangle = 0$,
$\langle \Psi^0_X| H_{em}|\Psi_G\rangle =
\frac{\lambda J_{sd}}{\sqrt{2}} \sum_I 
\Phi(\vec{R}_I) M^z_I$, 
$\langle \Psi^{+1}_X| H_{em}|\Psi_G\rangle = 
-\frac{\lambda J_{sd}}{2} \sum_I 
\Phi(\vec{R}_I) M^-_I$, and
$\langle \Psi^{-1}_X| H_{em}|\Psi_G\rangle = 
+\frac{\lambda J_{sd}}{2} \sum_I 
\Phi(\vec{R}_I) M^+_I$,  
where $\lambda=\alpha-\beta$ and 
$\Phi(\vec{R}_I)\equiv\varphi_+(\vec{R}_I)\varphi_-(\vec{R}_I)$.
We finally find
\begin{eqnarray}
H^{\rm eff}_{mm} = \sum_{I,I'}
\Delta_{I I'} \vec{M}_I \cdot \vec{M}_{I'},
\label{Heffmm}
\end{eqnarray} 
where
$\Delta_{I I'} = - \frac{\lambda^2 J^2_{sd}}{2 \Delta^{(e)}_{ST}}
\varphi_+(\vec{R}_I)\varphi_-(\vec{R}_I)
\varphi_+(\vec{R}_{I'})\varphi_-(\vec{R}_{I'})$.
Note that the Mn-Mn coupling 
for a lateral quantum dot molecule with two magnetic impurities 
localized at the center of each dot is given by
\begin{eqnarray}
\Delta_{12} = + \frac{\lambda^2 J^2_{sd}}{2 \Delta^{(e)}_{ST}}
\frac{\varphi^2_L(\vec{R}_1)\varphi^2_R(\vec{R}_2)}{4(1-W^2)} > 0.
\label{Delta_12}
\end{eqnarray} 
Here we assume that $\vec{R}_1$ and $\vec{R}_2$ are the position of Mn's 
centered at left and right dots, and therefore 
the electron wave-functions at the opposite position of Mn's,
$\varphi^2_L(\vec{R}_2)$ and $\varphi^2_R(\vec{R}_1)$ are negligible
due to high localization of the wave-functions. 
Because $\Delta_{12}$ is positive, 
the coupling between two Mn mediated by electrons is anti-ferromagnetic
with $M=0$ as the ground state.
For Mn, this state is separated by an
energy gap, $30\Delta_{12}$, from the ferromagnetic state $M=M_1+M_2(=5)$.
There are series of canted states with $M=1,\dots,4$ between $M=0$ and
$M=5$.

\section{Exact diagonalization}
The effect of $V_p$ at $B=0$ on both $\alpha$ and $\Delta^{(e)}_{ST}$ 
is shown in Fig. \ref{fig2}.
This calculation is based on CI method using 20 SP-orbitals (400 electronic 
configurations).
With increasing $V_p$ the inter dot tunneling and 
the overlap between L and R wave-functions decreases.
This results to the decrease of $\alpha$ 
(hence $\lambda$) and $\Delta^{(e)}_{ST}$ simultaneously. 
The expansion of the ground state wave function $|\Psi_G\rangle$,
in terms of leading configurations of two electrons is shown.
The contribution of the rest of configurations is negligible.

With increase of magnetic field, the electron spin singlet-triplet energy gap
$\Delta^{(e)}_{ST}$ decreases.
Close to the transition point, $\Delta^{(e)}_{ST}$ vanishes, and
the perturbation method 
fails. 
An unpertabative approach has to develop to calculate
the low lying energy states of $H$
in order to map
$H$ into  
$H^{\rm eff}_{mm}$. 
Here we exactly diagonalize 
Hamiltonian $H$ by expanding the many body wave-function in 
the basis of electron-Mn configurations: 
$|\Psi, M\rangle = c^\dagger_{\alpha\sigma} c^\dagger_{\beta\sigma'} |0 \rangle
\otimes |M_{z1},M_{z2}\rangle$.
Because of $[S_z,H_{em}]\neq 0$, 
($S$ is the total spin operator of two electrons), 
the states with different $S_z$ are mixed, 
hence the dimension of matrix $H$ that has to be diagonalized is 
given by  $N_C=(2M_1+1)(2M_2+1) \sum_{N_\uparrow=0}^N N_{SP}!
/[N_\uparrow! (N_{SP}-N_\uparrow)!] N_{SP}!/[N_\downarrow! 
(N_{SP}-N_\downarrow)!]$. 
$N=N_\uparrow + N_\downarrow$ is the number of electrons (here $N=2$),
and $N_{SP}$ is the number of single particle orbitals.
To check the convergence of CI we perform exact diagonalization 
using single particle orbitals up to $N_{SP}=20$. 
The result of this calculation and the magnetic phase diagram of Mn
is summarized in Fig. \ref{fig1} where
the electron spin singlet-triplet phase diagram is calculated by 
configuration interaction method 
for two electrons in lateral quantum
dot molecules. 

\begin{figure}
\begin{center}
\vspace{1cm}
\includegraphics[width=0.98\linewidth]{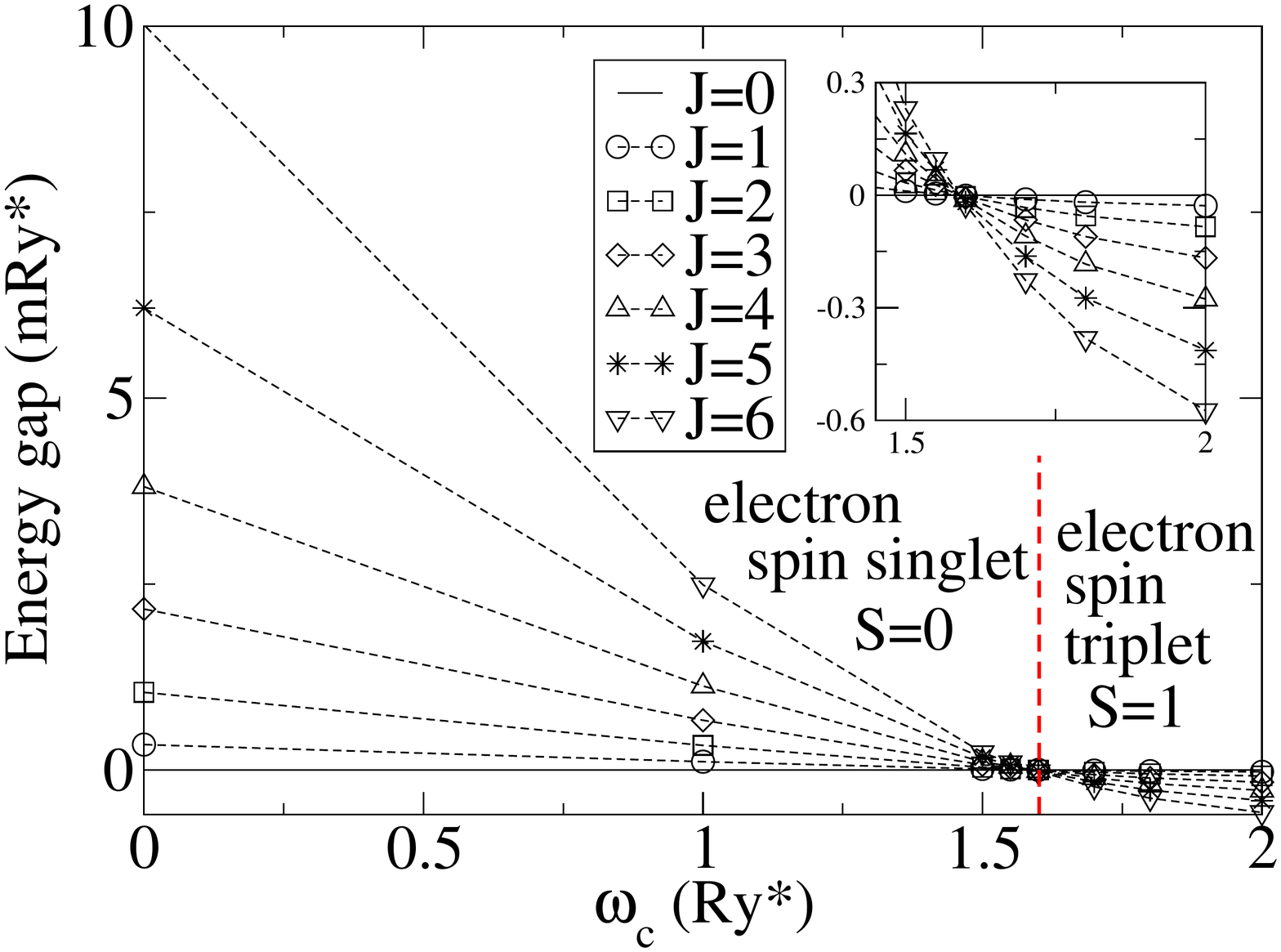}
\caption{
(Color online) 
Energy gap of a system of two electrons and two Mn in lateral quantum dot
molecule with $V_p=7$ as a function of cyclotron energy (magnetic field). 
At $\omega_c=1.55$ electron spin singlet-triplet transition 
is seen.
The vertical dashed line marks the transition point.   
This transition induce para-ortho transition in Mn's.
Below (above) this transition the ground state is identified by total
angular momentum $J=0$ ($J=6$). 
A magnified part of main figure is shown in inset.  
}
\label{fig3}
\end{center}
\end{figure}

Fig. \ref{fig3} shows the lowest energy gap,
$\Delta=E_J - E_{J=0}$, calculated for two electrons 
and two Mn in lateral quantum dot
molecule as a function of cyclotron frequency $\omega_c=eB/m^*c$ and $J$.
Here $\vec{J}=\vec{M}+\vec{S}$ is the total electron-Mn spin operator
($\vec{M}=\vec{M}_1+\vec{M}_2$, and $\vec{S}=\vec{S}_1+\vec{S}_2$
are total Mn and electron spin operators).
For illustration we switched off the electron and Mn Zeeman couplings.
Here the singlet-triplet transition occurs because of 
change in wave functions and e-e Coulomb matrix elements.
Close to the transition point where the single particle energy levels of
valence electrons are degenerate (half-filled),
the e-e Coulomb interaction leads
singlet-triplet transition in accordance with spin Hund's rule.
The eigenvalues of $H$ are grouped into $J=0,\dots,6$.
States with given $J$ are $2J+1$-fold degenerate.
It is convenient to characterize these states based on the total spin of 
electrons, e.g., spin singlet ($S=0$) and triplet ($S=1$) 
and total spin of two Mn with $M=0,\dots,5$.
In this work we are interested in the magnetic ordering of two Mn that can be 
described by 
anti-ferromagnetic, ferromagnetic and canted states corresponding to 
$M=0$, $M=5$, and $M=1,\dots , 4$.
As it is shown in Fig. \ref{fig3} 
spin of electrons undergo singlet-triplet transition  
at $\omega^*_c=1.55$ and $V_p=7$.
Within $\omega_c<\omega^*_c$, $J=M=S=0$ is 
the non-degenerate ground state.
At $\omega_c=\omega^*_c$, the energy gap of antiferromagnetic, 
ferromagnetic and canted states vanish all together and
the ground state switches to ferromagnetic
state with maximum spin multiplicity corresponding to 
$J=6$, $M=5$, and $S=1$.
In the limit of strong magnetic field there are $2J+1=13$ degenerate 
states that form the ground state.
However this degeneracy is removed by Zeeman coupling that 
guarantees the uniqueness of the ground state with $M_z=-5$ and $S_z=-1$.  
Within the resolution of our exact diagonalization we did not observe
any range of magnetic field that the ground state exhibits
canted ordering.
In Fig. \ref{fig3} at $B=0$ and $V_p=7$
we compare the energy gap calculated using CI, 
$E_{J=6} - E_{J=0} = 10$ mRy$^*$, with perturbation approach.
We find $\Delta^{(e)}_{ST}= 0.013$, $\Delta_{12}=0.38$, 
hence $E_{J=6} - E_{J=0} = 11.4$ all in mRy$^*$,
in qualitative agreement with exact energy gap.

\section{Conclusion}
In this work we studied phase diagram of quantum dot molecules consist 
of two electrons and two magnetic impurities (Mn) confined in each dot. 
We demonstrated that the spin singlet-triplet transition of two electrons 
that are controlled by external electric gate voltage and magnetic field, 
can induce ferromagnetic-antiferromagnetic transition in the magnetic 
impurity system. Therefore, Mn-Mn spin transitions mediated by e-Mn exchange 
interaction can be controlled indirectly by external electric gate voltage 
and magnetic field. This allows us to suggest application of spin of magnetic 
impurities for the entanglement of qubits in quantum information processing. 
The advantage of using spin of magnetic impurities as qubit, instead of QD 
electrons resides in the possibility in achieving higher spin coherence time. 
In analogous to the magnetic moment of nuclear impurities in host semiconductor
systems, we speculated that the spin coherence time in the Mn system is 
expected to be longer than the QD electron system due to small localization 
length of Mn d-orbitals that suppresses the qubit spin-orbit coupling 
as well as hyperfine interaction with the magnetic moment of nuclei of host 
semiconductor. Our analysis based on exact diagonalization allows mapping the
electron-electron and electron-Mn Hamiltonian $H$ into an 
effective Mn-Mn Heisenberg 
Hamiltonian $H^{\rm eff}_{mm} = \Delta_{12} \vec{M}_1\cdot\vec{M}_2$ 
in agreement with the perturbative results, e.g., an RKKY model 
calculated for weak coupling
at low magnetic fields.
Consistent with the magnetic field dependence of 
the lowest lying states of full Hamiltonian $H$, 
$\Delta_{12}$ changes sign at critical magnetic field that
leads to spin singlet-triplet transition of two electrons in lateral 
quantum dot molecules.
This is a level crossing that results to para-ortho transition 
induced by electrons in 
artificial H$_2$ molecules where the magnetic
impurities resembling the magnetic moment of 
nucleus of the actual H$_2$ molecules,
and 
the interaction between two Mn at high magnetic field is 
determined by the magnetic polaron effect.

\section{acknowledgment}
Author acknowledges partial support from 
the US ONR Grants N00014-06-1-0616, N00014-06-1-0123,
and NSF ECCS, and thanks Thomas Brabec, Sasha Govorov, Pawel Hawrylak,  
Miguel Angel Martin-Delgado,  
Andre Petukhov, Dean Sherry, and Igor Zutic
for comments and discussions.




\begin{thebibliography}{99}
\bibitem{comment}
Because the nuclear wave function of H$_2$ molecule is odd under exchange of 
two protons, in the gas phase, the para and ortho energy gap can be expressed 
as the energy difference between nuclear rotational energy levels associated 
with the molecular angular momentum.
In the solid phase, the rotational degree of freedom of H$_2$ molecules is 
frozen, and the para and ortho states of hydrogen are degenerate. 
The spin degeneracy is removed by the electron-proton hyperfine interaction
that is analogous to the e-Mn contact interaction in artificial H$_2$ quantum 
dot molecule.
\end{thebibliography}
\end{document}